# Temperature Dependence of Superconducting Gaps in Mg$_{1-x}$Al$_x$B$_2$ System Investigated by SnS-Andreev Spectroscopy

S. A. Kuzmichev[1*], T. E. Shanygina[1,2], S. N. Tchesnokov[1], S. I. Krasnosvobodtsev[2]

[1] *M.V. Lomonosov Moscow State University, 119991, Moscow, Russia*
[2] *P.N. Lebedev Physical Institute of the RAS, 119991, Moscow, Russia*

**Abstract**

Detailed temperature dependence of both superconducting gaps was obtained directly by means of SnS-Andreev spectroscopy. The $\Delta_{\sigma,\pi}(T)$-curves were shown to be deviated from standard BCS-like behavior, due to $k$-space proximity effect between $\sigma$- and $\pi$-condensates, which could give a key to experimental determination of interband electron-phonon coupling constants. For the first time, an excellent qualitative agreement with theoretical predictions of Nicol and Carbotte, and Moskalenko and Suhl was shown. dI(V)/dV-spectra of SnS-Andreev contacts based on $MgB_2$ samples (with defects of crystal structure), and $Mg_{1-x}Al_xB_2$ polycrystalline samples (with the local critical temperature $T_C$ variation $10\,K \leq T_C \leq 37\,K$) were studied by means of the "break-junction" technique within the temperature range $4.2\,K \leq T \leq T_C$.

A binary superconducting compound $MgB_2$ with enormously high critical temperature $T_C = 40\,K$ [1] remains to be of research interest so far. However, the pioneer theoretical works studied two-gap superconductors [2] have appeared as far back as the middle of the last century. Theoretical calculations [3–10] and some experimental studies [11–23, reviews 24, 25 and refs. therein] have ascertained an existence of two superconducting gaps at the different sheets of Fermi surface [4,26]: the large two-dimensional (2D) hole condensate described by $\Delta_\sigma = 7 \div 12$ meV gap and the small 3D one with $\Delta_\pi = 2 \div 0.5$ meV. According to the theory elaborated by Choi *et al.* [27] $MgB_2$ quasiparticle density of states is characterized by two well-legible split gap peculiarities related to the four different bands including two $\sigma$- and two $\pi$-bands (the four-band model) that can be considered as two effective ones (the two-band approach) [28]. The rather high $T_C$ value originates from $\sigma$-bands coupling on the $E_{2g}$ boron phonon modes [5]. A large boron isotope effect [29] gives support unambiguously to phonon-mediated superconductivity [30].

---

[*] Corresponding author. Tel.: +7(495)939-39-41; fax: +7(495)932-88-76
E-mail: kuzmichev *at* mig.phys.msu.ru

Impurity injection strongly affects the critical temperature of magnesium diboride. The lattice parameter $c$ and the electron-phonon coupling constants $\lambda_{ij}$ value lowering due to Al concentration $x$ increasing [31] lead to the $T_C$ diminution to zero at $x \to 0.5$ in $Mg_{1-x}Al_xB_2$ [32,33]. Theoretical calculations of Ummarino *et al.* [31] indicated $Mg_{1-x}Al_xB_2$ to turn into the dirty limit at $19K \leq T_C \leq 25K$. Later, Kortus *et al.* [34] shifted the gaps merging downward the lower $T_C = 10 \div 15K$. At the same time, the available experimental studies on the dirty limit transition are still far from completion. The only gap existence with $2\Delta/kT_C < 3.52$ reported initially [20,21] seemed to be not convincing and was not confirmed later: the authors [22] observed two distinct gaps till $T_C \approx 15\,K$ both in single-crystal and polycrystalline $Mg_{1-x}Al_xB_2$. Nevertheless, when approximating the experimental data, Klein *et al.* [35] suggested the dirty limit transition at $T_C \approx 13\,K$, which seemed doubtful for the reason of $2\Delta/kT_C$ given above. Such a contradiction between the experimental data [20-22,35] was attributed to the interband scattering value varying in dependence of the experimental technique [34]. Besides, by authors' own admission [36], a problem of detecting the large gap peculiarities at SN-junctions conductance curves and a need in data fitting using up to 7 adjustment parameters leads to rather high uncertainty in superconducting gaps definition.

The present work implies comprehensive experimental study of more than 120 current-voltage characteristics (CVCs) and their derivatives of Sharvin-type SnS-contacts based on $MgB_2$ and $Mg_{1-x}Al_xB_2$ within the temperature range $4.2\,K \leq T \leq T_C$ as well as *directly* obtained high-quality gaps temperature dependence.

Three sets of polycrystalline magnesium diboride samples were exploited: Kr-series (with excess of Mg and transition width $\Delta T_C \cong 0.3\,K$; bulk $T_C$ up to 41.5 $K$ determined from the dR(T)/dT maximum [37]); MB-series (the tablets have been pressed from $MgB_2$-powder with structural disorder containing up to ~ 20% MgO impurity, with the range of bulk critical temperatures $20\,K \leq T_C \leq 40\,K$), and MBA-series ($Mg_{1-x}Al_xB_2$ with the range of aluminum concentrations $0.15 \leq x \leq 0.45$ and bulk $10\,K \leq T_C \leq 37\,K$ [38]).

We used Andreev spectroscopy and intrinsic Andreev spectroscopy (intrinsic multiple Andreev reflections effect (IMARE) that usually exists due to presence of steps-and-terraces at clean cryogenic clefts) for determining superconducting properties of $Mg_{1-x}Al_xB_2$. Andreev nanocontacts in polycrystalline samples were prepared by means of the "break-junction" technique [39,40]. Our samples represented thin plates with the dimensions $\cong 1.5 \times 0.5 \times 0.2\,mm^3$ have been attached with two current and two potential leads by liquid In-Ga alloy. A microcrack into the samples was generated by bending a sample holder at helium temperature, therefore, we dealt with mechanical contact of two clean cryogenically cleaved surfaces separated to a moderate distance, thus excluding impurity affecting on cryogenic clefts. I(V)-, dI(V)/dV-characteristics and temperature dependences of resistance were measured on a digital set-up with the standard modulation technique [41,42] at Moscow State University and University of Wuppertal.

A subharmonic gap structure (SGS) appears at SnS-contact dI(V)/dV-characteristics due to multiple Andreev reflections effect [43]. In the case of two-band superconductor, it shows sets of dynamic conductance dips at bias voltages [44]:

$$V_n = \frac{2\Delta_{1,2}}{en}, \text{ where } n = 1, 2, 3, etc. \qquad (1)$$

To get accurate values of $\Delta_{1,2}$ one need to plot SGS minima positions $V_n$ as a function of $1/n$; this line should start from (0, 0) point. In accordance with [44] the dynamic conductance dips' amplitude decreases while $n$ increases. This fact helps to separate the SGS caused by the large gap from the SGS of the small gap in two-band superconductors, since an amplitude of the first minima that corresponds to the small gap ($n_2 = 1$) is usually much higher than dips related to $n_1 = 3,4$ of large gap. Thus, one can easily detect a range of both SGS.

On using the "break-junction" technique, the layered structure of $Mg_{1-x}Al_xB_2$ allows one to get so-called stacks of Andreev contacts on cryogenic clefts, i.e. IMARE on junctions of S-n-…-S-n-S-type that was observed earlier in Bi-2201 [45]. The bias voltage of any peculiarities caused by the bulk properties at dI(V)/dV-characteristic of such an SnS-array scales with the number of consistently connected contacts $N$ in comparison with single junction's one. On the contrary, this is not a case for the location of CVC features that demonstrate a surface defect influence. One could determine $N$ by collating dynamic conductance spectra of Andreev arrays with different numbers of junctions in a stack and normalizing them to a single contact characteristics. As the array CVCs are nearly free of surface affecting, the measuring of such characteristics provides information on bulk superconducting properties and helps with accurate gap values definition. In previous investigations of our group [46-48] we have ascertained $\Delta_\sigma = 11 \pm 1$ meV and $\Delta_\pi = 2 \pm 0.5$ meV for $MgB_2$ (with maximal $T_C = 40 \pm 1 K$).

The sharpest SGS is usually observed only in qualitative Andreev contacts of a diameter less than quasiparticle mean free path. The resistance of a classical Sharvin contact (ballistic limit) [49] can be defined as

$$R = \frac{4}{3}\pi\left(\frac{\rho l}{a^2}\right), \qquad (2)$$

The product of the normal state bulk resistivity and quasiparticle mean free path was shown to be of the order of $\rho l = p_F / ne^2 \cong 2 \cdot 10^{-12} \Omega \cdot cm^2$ [50] (where $p_F$ is a Fermi impulse), so that taking into account an average resistance of our contacts $R \sim 1 \div 10\,\Omega$ one could obtain Sharvin contact's diameter $a \sim 0.01 \div 0.03\,\mu m$ that is smaller than the samples' crystal grain diameter.

Note that SnS-Andreev spectroscopy is a convenient tool for superconducting gaps' temperature dependence measurements, because SGSs at characteristics of symmetrical SnS-contacts remain rather well-distinguished right up the *local* $T_C$ of the contact (temperature of contact transition to a normal state). Coinciding of local $T_C$ of the contact and bulk $T_C$ of the whole sample is observed only in the most clean and homogeneous samples. To define gap values *no fitting* of conductance curves is required as in the case of SN-Andreev spectroscopy [44]: at any temperatures up to $T_C$ they can be obtained directly using formula (1).

*Fig. 1(a)* represents normalized to a single contact dI(V)/dV-characteristic of four junction SnS-array (contact Kr10c, Kr-series of $MgB_2$ samples) measured within the temperature range $4.2 K \leq T \leq T_C$. The well-visible minima reflect two subharmonic gap structures (marked by black and gray labels). In accordance with the theory of Kümmel *et al.*

[44] at any temperatures less than $T_C$ SGS dips' voltage $V_n$ plotted versus $1/n$ forms a line dependence (see formula (1)). The functions for both SGSs at $T \approx 4.2\ K$ are drawn up in an inset of *Fig. 1(a)* and they give the gap values $\Delta_\sigma = 10$ meV and $\Delta_\pi = 2.4$ meV. Minima standing inside $n_\sigma = 3$ (marked by gray vertical dashes) correspond to $n_\pi = 1$, i.e. they start SGS for $\Delta_\pi$. This bias voltage (about 5 mV) also matches Andreev peculiarity $n_\sigma = 4$; in case it belongs to SGS of $\Delta_\sigma$, its amplitude must be less but not more than minima $n_\sigma = 3$. *Fig. 1(a)* shows the opposite. The latter and the temperature dependence deviation of minimum at 5 mV from BCS-like behavior give us a proof that this peculiarity shows position of $n_\pi = 1$. When the temperature increases to 11.5 $K$, the $\pi$-gap starts to close while the $\sigma$-gap remains unchanged; this leads to an appearance of $n_\sigma = 4$ at dI(V)/dV-characteristics. As regards the $n_\sigma = 1$ region, the minima are weak and widely distributed. Such an $n_\sigma = 1$ behavior is reproduced from time to time for the samples of Kr series that have an excess of Mg; the reasons for it are not quite clear.

The taken value of *local* $T_C \approx 41.6\ K$ conforms to temperature of dI(V)/dV-characteristic becoming horizontal. Knowing of the local $T_C$ for each contact can facilitate in BCS-ratios for both gaps' accurate calculation. According to our data, the BCS ratio is $2\Delta_\sigma / kT_C \approx 5.6$ here, while $2\Delta_\pi / kT_C \approx 1 << 3.52$ testifies to $\pi$-band superconductivity proximity-induced by $\sigma$-condensate in the *k*-space within the range $T > T_C^\pi$ (where $T_C^\pi$ is a critical temperature defined by only $\pi$-bands' properties without interaction with $\sigma$-condensate and has been of the order of 15 K [6] at the optimal doping level). At lower $T_C$ the $\pi$-gap BCS-ratio value tends to the BCS-limit 3.52 [46]. The latter evidently proves 3D $\pi$-condensate superconductivity to be characterized by BCS-like coupling.

Using data of *Fig. 1(a)* the temperature dependence of superconducting gaps has been derived (see *Fig. 1(b)*) *directly*, i.e. without any additional fittings. The curves significantly differ from each other, and both of them deviate from the standard single-gap BCS-course. Small gap tends to close at its "own" $T_C^\pi$ then proceeds slowly fading, like a "tail", to the common $T_C$ of the junction. At the same time, $\Delta_\sigma(T)$-curve slightly bends at temperatures $T \sim T_C^\pi$, thus passing below the single-gap BCS-like dependence (dash-dotted line).

Therefore, the observed common $T_C$ of contact slightly decreases in comparison with the "own" $T_C^\sigma$ of $\sigma$-condensate [6].

Nicol and Carbotte, based on Moskalenko's and Suhl's system of gap equations [2], have shown [51] that the shape of gap temperature courses can vary in dependence of interband and intraband electron-phonon coupling constants $\lambda_{ij}$ set. In particular, in the case of $\lambda_{\sigma\pi} = \lambda_{\pi\sigma} = 0$ the two gaps are independent, with the small gap closing at its own $T_C^\pi$ (see inset of *Fig. 1(b)*, solid line). In the case of nonzero interband interaction gaps temperature dependence characterized by two different $\lambda_{ij}$ sets is also depicted at the inset of *Fig. 1(b)* (dashed and dotted lines). Our experimental $\Delta_\sigma(T)$ and $\Delta_\pi(T)$-curve shapes are in good accordance with one of theoretical cases considered by Nicol and Carbotte [51].

*Figure 2(a)* shows dI(V)/dV-characteristics of the two junction SnS-Andreev stack on the $Mg_{1-x}Al_xB_2$ sample cleft (contact MBA3t, MBA-series). Andreev reflection minima bias voltages are marked by black labels ($\sigma$-gap) and gray labels ($\pi$-gap). Despite the relatively low value of *local* $T_C = 13.5 \pm 1\ K$, sharp SGSs are observed at the whole range of temperatures measured. $\Delta_\sigma(T)$ and $\Delta_\pi(T)$ curves derived using data of *Fig. 2(a)* are depicted in *Fig. 2(b)*; their outlook is similar to the previous plot in *Fig. 1(b)*. According to results of the theoretical work [52] symmetry of $\Delta_\sigma$-minima (as one can see in *Figs. 1(a)* and *2(a)*) points to the s-wave type of $\sigma$-gap order parameter.

Superconducting gaps' temperature dependence for $MgB_2$ and $Mg_{1-x}Al_xB_2$ SnS-Andreev contacts with different local $T_C$ is presented in *Fig. 3* for comparison. The values of $\Delta_\pi(0)$ are reproduced for both stack and single contacts. The latter means that $\Delta_\pi$-gap is an intrinsic property of $Mg_{1-x}Al_xB_2$, as well as the small gap dependences obtained could not be interpreted as a surface superconducting gap ones. It is clearly seen that $\Delta_\sigma(T)$ and $\Delta_\pi(T)$ accents, like appearance of a "tail" at the $\pi$-gap course and a bend at the $\sigma$-gap one, are well-reproduced within the whole range of local critical temperatures. This fact points to a slight change in the interband interaction and the $\lambda_{\pi\sigma}/\lambda_{\sigma\pi}$-ratio while either doping electrons amount increase in boron planes or the extent of disorder increases. We wish to note that smooth BCS-ratio changing from 5.8 to 7 for the $\sigma$-gap with decreasing of $T_C$ in *Fig. 3* is accidental. The summary result of hundreds of $MgB_2$ and $Mg_{1-x}Al_xB_2$ contacts investigation done with the authors' participation has shown that BCS-ratio value is about $6\pm1$ for all groups of the samples in the doping range [47]. Relatively low bend rates at $\Delta_\sigma(T)$ suggest the weak interband coupling. The dirty limit transition was not observed till the local $T_C \approx 11\,K$.

Based on the two-band Eliashberg model [53], subsistent theoretical studies [7-9, 27, 28, 51] yield rather low BCS-ratio $2\Delta_\sigma/kT_C = 4 \div 4.6$ and, therefore, cannot entirely explain superconducting properties of $Mg_{1-x}Al_xB_2$ system, in particular, the experimental $2\Delta_\sigma/kT_C = 5 \div 7$ [13, 16, 46-48]. Our data support the theoretical calculations for $\Delta_\sigma(T)$ and $\Delta_\pi(T)$ presented in [51]. We have not observed single-gap BCS-like $\Delta_i(T)$-dependences predicted in [7-9,27,28,31] which points to overstated value of interband (nondiagonal) coupling constants and low ratio $\lambda_{\pi\sigma}/\lambda_{\sigma\pi} \sim 1.35$ calculated in [7,27,28]. Besides, Leggett plasma modes [54], first found out in the previous works [46,48,55], would be unobservable, taking into account the large $\lambda_{ij}$ values from these theoretical works. Such nontrivial behavior of $\Delta_\sigma(T)$ and $\Delta_\pi(T)$-curves supposed to be a consequence of $k$-space proximity effect between two superconducting condensates [56] due to nonzero interband interaction.

In conclusion, the current-voltage characteristics and their derivatives of $MgB_2$ and $Mg_{1-x}Al_xB_2$ SnS-Andreev contacts were investigated within the temperature range $4.2\,K \leq T \leq T_C$. The directly obtained temperature dependences $\Delta_{\sigma,\pi}(T)$ differ from standard single-gap BCS-course and from each other: appearance of the "tail" at $\Delta_\pi(T)$ and the bend at $\Delta_\sigma(T)$ testifies to an influence of the $\sigma$-condensate to the $\pi$-one ($k$-space proximity effect). The data were shown to be in excellent qualitative agreement with theoretical predictions of Nicol and Carbotte [51], and Moskalenko and Suhl [2].

The authors are grateful to Prof. Ya.G. Ponomarev for helpful discussions, providing techniques and materials, as well as L.G. Sevastyanova, K.P. Burdina, V.K. Gentchel, B.M. Bulychev for sample synthesis and H. Piel for an opportunity to use "break-junction" set-up at University of Wuppertal. We also thank V.A. Stepanov for fruitful debates.

**Figure captions:**

**Fig. 1. (a)** SGS at the normalized dI(V)/dV-characteristic of *MgB$_2$* SnS-stack (four junctions) within the temperature range $4.2\,K \leq T \leq T_C$ (local $T_C = 41.6\,K$, Kr-series, contact Kr10c). Andreev minima caused by $\Delta_\sigma$ and $\Delta_\pi$ are marked by black and gray labels correspondingly. The curves were shifted along the vertical scale for convenience. **(Inset)** Linear SGS dips' voltage $V_n$ $(1/n)$ dependence for both gaps. The dashed lines slope gives the gap values $\Delta_\sigma = 10 \pm 0.3$ meV and $\Delta_\pi = 2.4 \pm 0.3$ meV.

**(b)** The temperature dependence of superconducting gaps $\Delta_\sigma$ (black circles) and $\Delta_\pi$ (black empty circles) for MgB$_2$ SnS-contact with one of the highest local T$_C$ plotted under the data of *Fig. 1(a)*, as well as the bulk resistance dependence for this sample (light gray empty circles connected by line). Standard single-gap BCS-like behavior (black dash-dot lines) is presented for comparison. **(Inset)** Theoretical $2\Delta_{\sigma,\pi} / kT_C$-dependences of MgB$_2$ for $\lambda_{\sigma\sigma} = 1$, $\lambda_{\pi\pi} = 0.5$ while nondiagonal values $\lambda_{\sigma\pi} = 0.1$, $\lambda_{\pi\sigma} = 0.01$ (dashed lines); $\lambda_{\sigma\pi} = 0.01$, $\lambda_{\pi\sigma} = 0.1$ (dotted lines) and $\lambda_{\sigma\pi} = \lambda_{\pi\sigma} = 0$ (solid line) from ref. [51].

**Fig. 2. (a)** SGS at the normalized dI(V)/dV-characteristics of Mg$_{1-x}$Al$_x$B$_2$ SnS-stack (two junctions) within the temperature range $4.2\,K \leq T \leq T_C$ (local $T_C \approx 13.5\,K$, MBA-series, contact MBA3t). Locations of Andreev minima describing $\Delta_\sigma \approx 4.2$ meV and $\Delta_\pi \approx 1.4$ meV are marked by black and gray labels correspondingly. The curves were shifted along the vertical scale for convenience.

**(b)** The temperature dependence of superconducting gaps $\Delta_\sigma$ (circles) and $\Delta_\pi$ (open circles) for the Mg$_{1-x}$Al$_x$B$_2$ SnS-contact plotted under the data of *Fig. 2(a)*. Standard single-gap BCS-like behavior (black dash-dot lines) is presented for comparison.

**Fig. 3.** The temperature dependence of σ-gaps (dark symbols) and π-gaps (light symbols) obtained on MgB$_2$ contacts: Kr10b (gaps value: $\Delta_\sigma \approx 10$ meV, $\Delta_\pi \approx 1.7$ meV, up triangles), MB6 ($\Delta_\sigma \approx 7.6$ meV, $\Delta_\pi \approx 2$ meV, rhombs), MB12 ($\Delta_\sigma \approx 6.1$ meV, $\Delta_\pi \approx 1.7$ meV, squares), as well as Mg$_{1-x}$Al$_x$B$_2$ ones: contact MBA3t ($\Delta_\sigma \approx 4.2$ meV, $\Delta_\pi \approx 1.4$ meV, circles), and MBA3a (**Inset**; $\Delta_\sigma \approx 3.1$ meV, $\Delta_\pi \approx 1.7$ meV, down triangles). Single-gap BCS-like curves (black dash-dot lines) are presented for comparison.

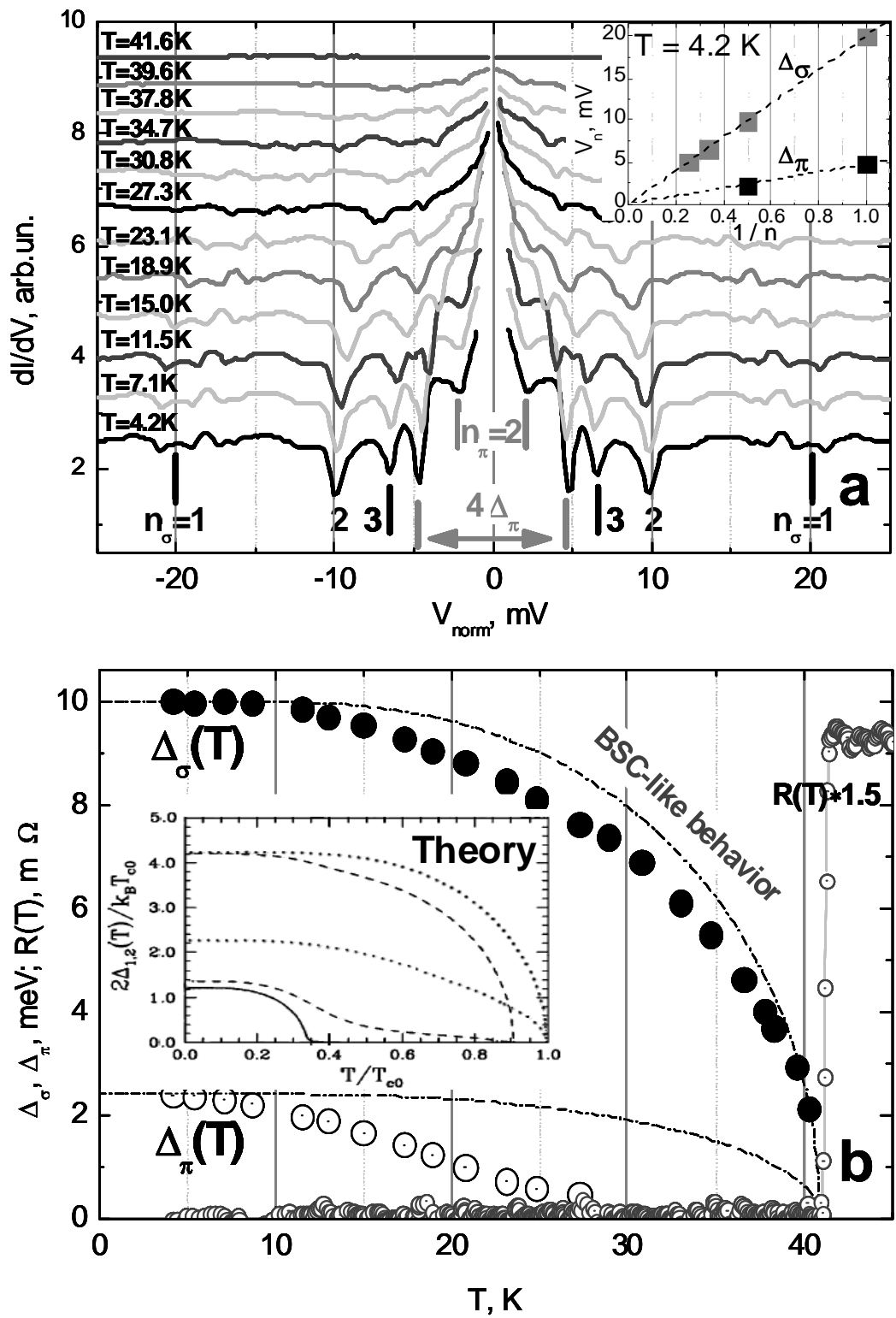

Fig. 1

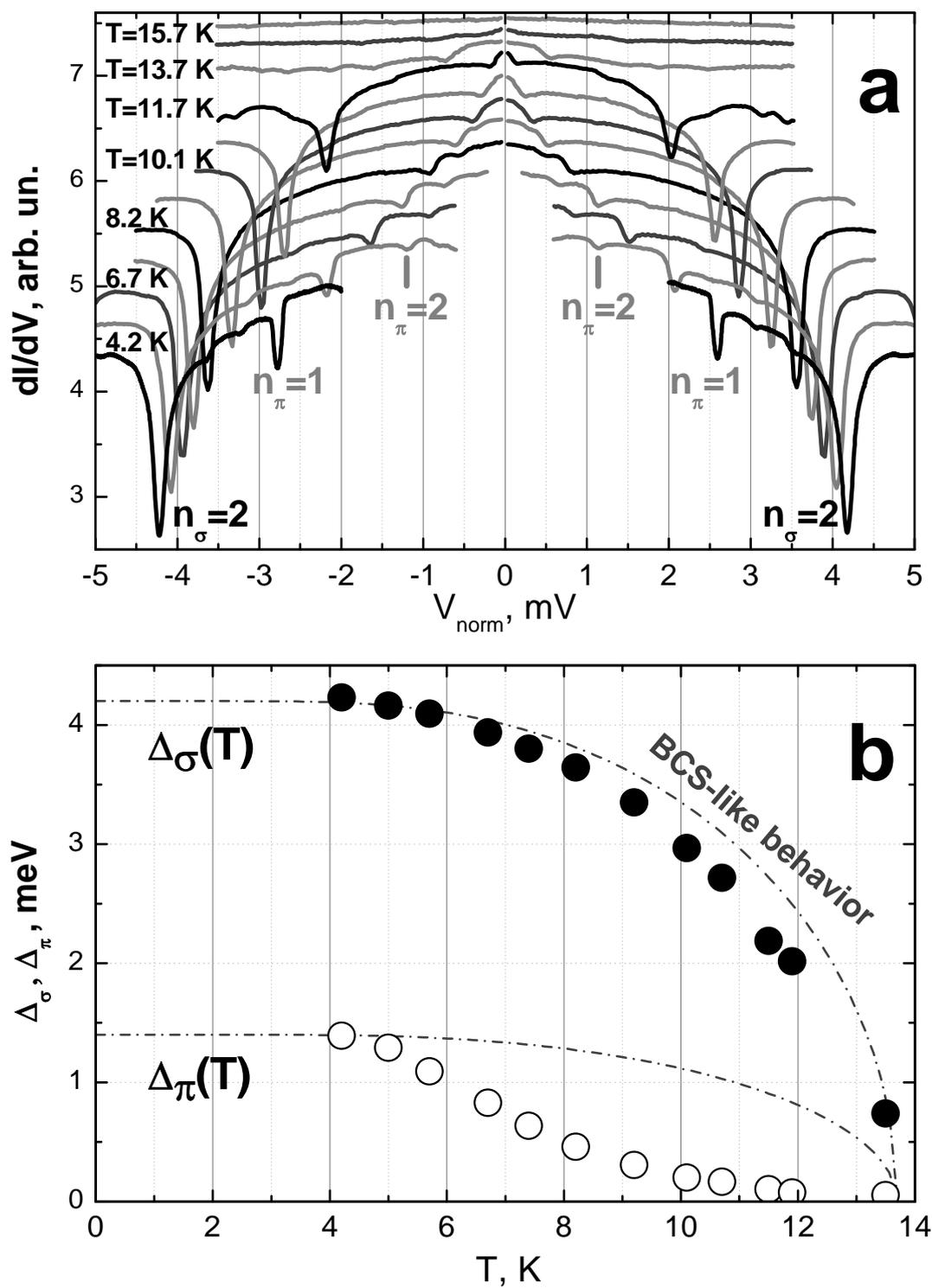

**Fig. 2**

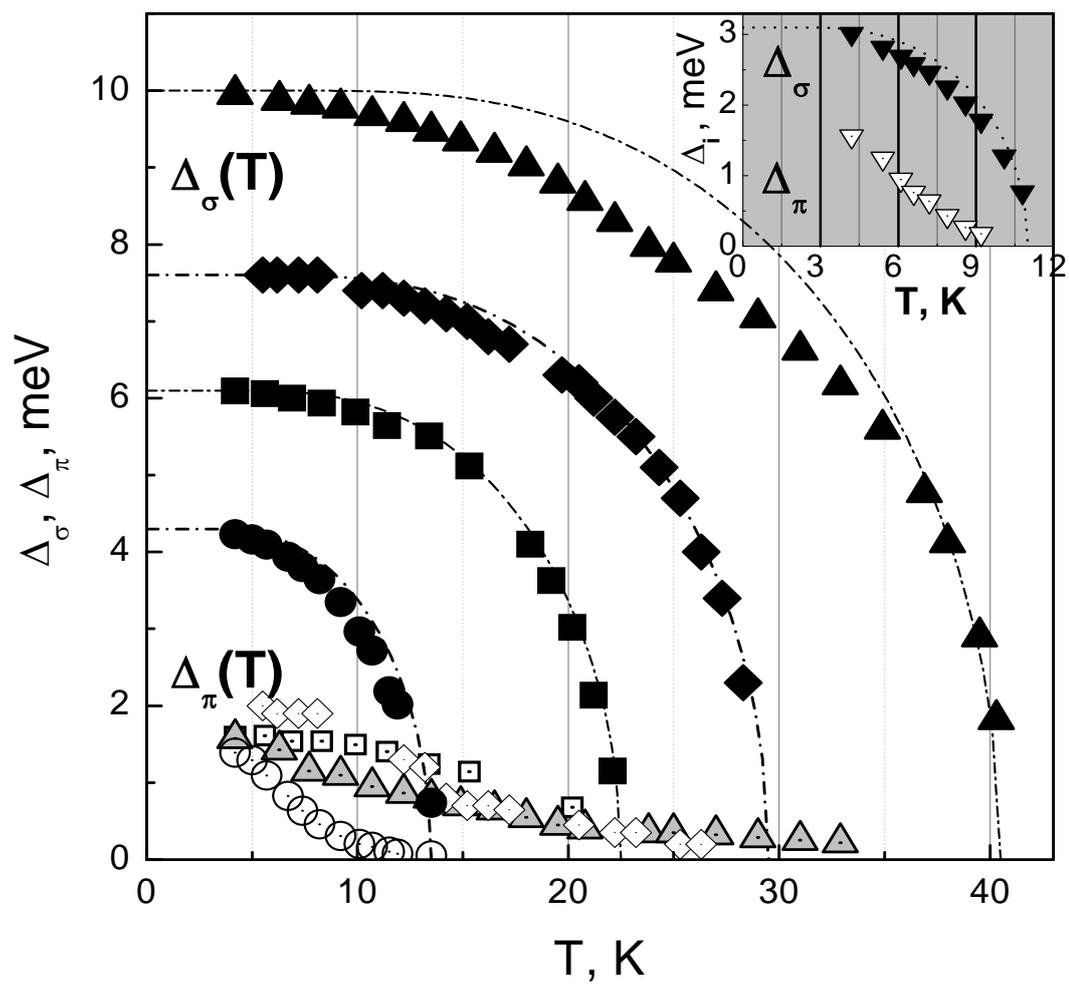

**Fig. 3**